# *Direct and inverse spin Hall effect: Lorentz force and Zeeman energy*

*A. Hernando[a,b,c], F. Gálvez[a,\*] and F. Guinea[b,c,d]*

a Instituto de Magnetismo Aplicado, UCM-CSIC-ADIF, P. O. Box 155, 28230-Las Rozas. Madrid, Spain and Dpto Física de Materiales, UCM, Ciudad Universitaria, 28040 Madrid, Spain
b IMDEA Nanociencia, Faraday 9, 28049 Madrid, Spain
c Donostia International Physics Center DIPC, Paseo Manuel de Lardizabal 4, 2018 Donostia-San Sebastián, Spain
d School of Physics and Astronomy, University of Manchester, Manchester, M13 9PY, UK
*Corresponding author: fgalvez@ucm.es

## *Abstract*

*It is shown that magnetic forces as the Lorentz force, exerted on electric currents, and the force $\mp\mu\nabla B$, exerted on electron spins at rest, account for both the transverse spin imbalance typical of spin Hall effect and the transverse charge imbalance associated with pure spin currents (inverse spin Hall effect). Considering that for stationary currents the laboratory reference frame and those for which the spin up and spin down carriers are at rest are inertial systems, one can easily find the forces exerted by the lattice on both spin sub-bands, as well as the force between sub-bands.*

The main idea underlying this article is based on the fact that the magnetic moment associated with the spin of the electric current carriers are subject to a force generated by the motion of the lattice in the opposite direction. This force is the same considered in the atomic spin-orbit coupling with the only difference that in this case the orbital current becomes a linear current. Provided that the whole system is electrically neutral, the current seen at the laboratory frame of reference, $S_BF$, due to the carrier motion, is the same that the current seen at the frame in which the carriers are at rest, $SF$, that is due to the relative lattice motion. The corresponding force exerted on the spins at *SF* can be obtained from the Zeeman energy as $\mp\mu\nabla B$, $\mu$ being the spin magnetic moment, the different sign corresponding to different spin polarization along the direction defined by the local magnetic field, and *B* is the inhomogeneous magnetic field seen in *SF* that is produced by the relative lattice current. Under this assumption it has been shown [1] that the force $\mp\mu\nabla B$ induces a transverse spin imbalance similar to that known as spin Hall effect [2] and with strength of the order of magnitude of that reported for spin Hall experiments with polarized conduction band [1,2,7]. In particular spin accumulation and Hall angle were shown [1] to be of the order of magnitude of that predicted by assuming intrinsic or extrinsic impurities and atomic spin-orbit scattering [3-7]. However, it seems at first sight that the lack of overall electrical current in a pure spin current makes impossible to explain the inverse spin Hall effect by effect of the induced magnetic fields.

In this article it is shown that magnetic forces account for the so-called inverse spin Hall [5], even though the net electric current in $S_BF$ is zero. Consider a thin metallic film in which along its longitudinal direction, *y*, flow two steady density currents $j_\uparrow = j_\uparrow n_\uparrow e v_\uparrow$ and $j_\downarrow = n_\downarrow e v_\downarrow$ corresponding to electron with spin up and down, respectively (see Figure 1). The spin sub-bands are polarized along the film thickness direction, *z*, that is determined by the local magnetic field produced by the currents [1]. The electron concentration is given by $n = n_\uparrow + n_\downarrow$. These electric currents generate magnetic forces, $F_{\uparrow\downarrow}$, between the assembly of electrons with spin up, $S_\uparrow$, and down, $S_\downarrow$, and forces exerted by the relative current of the positive background on both systems $F_{\uparrow B}$ and $F_{\downarrow B}$. All the forces act along the *x* direction since the magnetic moments or the magnetization lie along the *z*- axis and the currents are oriented along the *y*-axis

The force acting on a spin up electron, $f_\uparrow$, can be decomposed in two components: the force due to the assembly of spin down, $f_{\uparrow\downarrow}$, plus the force exerted by the positive background $f_{\uparrow B}$. The same decomposition holds for the spin down electrons.

The net force acting on the total spin system $F^H$, i.e. on $S_\uparrow + S_\downarrow$, can then be expressed as

$$F^H = n_\uparrow f_\uparrow + n_\downarrow f_\downarrow = n_\uparrow(f_{\uparrow\downarrow} + f_{\uparrow B}) + n_\downarrow(f_{\downarrow\uparrow} + f_{\downarrow B}) \quad (1)$$

That, after considering $n_\uparrow f_{\uparrow\downarrow} = -n_\downarrow f_{\downarrow\uparrow}$ leads to

$$F^H = n_\uparrow f_{\uparrow B} + n_\downarrow f_{\downarrow B} \quad (2)$$

This force pushes the electrons of both sub-bands along the *x* direction giving rise to a transverse charge imbalance. The difference between the forces acting on both sub-bands $F^S$ is the force inducing transverse spin separation along the same *x* direction and can be written as

$$F^S = (n_\uparrow f_{\uparrow B} - n_\downarrow f_{\downarrow B}) + 2n_\uparrow f_{\uparrow\downarrow} \quad (3)$$

To calculate the force components (see supplementary information a)), one can consider the reference frame $S_\uparrow F$ in which the spin up electrons are at rest. In this system, the force exerted by the $n_\uparrow$ electrons on the background density current of positive charges, $nev_\uparrow$, is the Lorentz force (Hall force) with modulus $nev_\uparrow\ n_\uparrow\mu_0\mu_B = nev_\uparrow\mu_0 M_s^\uparrow$, that acts along the x axis. Consequently, the force exerted by the background current on $S_\uparrow$ should be

$$n_\uparrow f_{\uparrow_B} = -nev_\uparrow\mu_0 M_s^\uparrow \quad (4)$$

By the same consideration one can find the force exerted on the $S_\downarrow$ sub-band to be $n_\downarrow f_{\downarrow_B}= -nev_\downarrow\mu_0 M_s^\downarrow$. By taking into consideration that $-\mu_0 n_\downarrow \mu_B = \mu_0 M_s^\downarrow$ and $\mu_0 n_\uparrow \mu_B = \mu_0 M_s^\uparrow$, the resultant force $F^H$ becomes

$$F^H = -\mu_0 M_s (j_\uparrow - j_\downarrow) \quad (5)$$

Where $M_s$ holds for the saturation magnetization of the whole conduction band $M_s = n\mu_B$

To calculate the force exerted by $S_\uparrow$ on the system $S_\downarrow$ we consider the Lorentz force exerted by the $n_\uparrow$ electrons on the relative density current $-n_\downarrow e(v_\uparrow - v_\uparrow)$ of negative charges (the electrons of $S_\downarrow$), seen also at $S_\uparrow F$. Then, this force becomes (see supplementary information b))

$$n_\uparrow f_{\uparrow\downarrow} = -n_\downarrow f_{\downarrow\uparrow} = -\mu_0 M(j_\uparrow - j_\downarrow) \quad (6)$$

where $M = (n_\uparrow - n_\downarrow)\mu_B$, is the overall magnetization of the conduction band

The force inducing spin separation is, according to eq. (3) and (6)

$$F^S = -\mu_0 M_s(j_\uparrow + j_\downarrow) + \mu_0 M(j_\uparrow - j_\downarrow) \quad (7)$$

The final results, relationships (5) and (7), can be also obtained in the laboratory reference frame, $S_BF$, as well as in the $S_\downarrow F$ frame since all of them are inertial reference frames.

The generality of these results can be illustrated after considering the following different scenarios.

1) *Spin Hall effect*. The film that is spin depolarized is connected to a battery that promotes an electric current *j*, as it is shown in Figure 2. In this case:
   $n_\uparrow = n_\downarrow = \frac{n}{2}$ and $v_\uparrow = v_\downarrow$ ; consequently $j_\uparrow = j_\downarrow = \frac{j}{2}$

The net force acting on the whole spin system according to eq. (5) is zero, the internal force $n_\uparrow f_{\uparrow\downarrow}$ is also zero, However, the absolute value of the difference between the forces exerted on the two sub-bands by the background is according to eq. (7)

$$F^S = \frac{n}{2}(f_{\uparrow_B} - f_{\downarrow_B}) = \mu_0 M_s j \quad (8)$$

This force induces spin imbalance known as spin Hall effect [2], since is opposite for the two sub-bands.

2) *Inverse spin Hall effect*. Consider now a film, also spin depolarized, in which there is a constant spin gradient between its extremes as shown in Figure 3.

As the conduction band is spin depolarized $n_\uparrow = n_\downarrow = \frac{n}{2}$. At the steady state there is pure spin current with $j_\uparrow = -j_\downarrow$ with strength proportional to the gradient. Then, according to eq. (5), there should appear a charge imbalance induced by $F^H = -\mu_0 M_s j$ or an effective Hall field

$$E^H = \frac{2\mu_0 M_s j_\uparrow}{ne} = \frac{\mu_0 M_s j}{ne} \qquad (9)$$

This field induce a Hall voltage known as inverse spin Hall effect. It is remarkable that a null net electric current induces Lorentz forces. The reason is that if the net current is zero is because we add two opposite real currents, each of them transported by carriers of opposite spin. In this case the background current, as well as its corresponding magnetic fields, are opposite to each other at $S_\uparrow$ and $S_\downarrow$. Then, the corresponding generated forces, the spins being also opposite, act on the same direction in both sub-bands, $S_\uparrow$ and $S_\downarrow$, thereby contributing to the charge imbalance given by (5)

3) *Full polarized conduction band.* In this case $n_\uparrow = n$; $n_\downarrow = 0$ and $j_\uparrow = j$

For the spin Hall effect, we find from eqs. (5) and (7)

$$F^H = F^S = M_s j \qquad (10)$$

The inverse spin Hall effect is the same that the spin Hall effect. The spin gradient induces the same current that a battery

$$F^S = F^H = M_s j \qquad (11).$$

4) *Spin Hall in a partially polarized band* (i.e. $n_\uparrow \neq n_\downarrow$. In this case $j_\uparrow = \frac{n_\uparrow}{n_\downarrow} j_\downarrow$. Then the charge imbalance field is not zero but according to (5) takes the value

$$F^H = -\mu_0 M_s j_\uparrow \left(1 - \frac{n_\uparrow}{n_\downarrow}\right) \qquad (12)$$

And the spin separating force, according to (7), becomes

$$F^S = = -\mu_0 M_s j_\uparrow \left(1 + \frac{n_\uparrow}{n_\downarrow}\right) \qquad (13)$$

In summary, it has been shown that the Lorentz force acting on electric currents (Hall effect) and the Zeeman force $\mp \mu \nabla B$ acting on the spin carriers, at the reference frames at which they are at rest, account for the properties and characteristics of the spin and inverse spin Hall effect. In fact, the results shown in this article indicates that both spin and charge imbalance originated by electric currents can be naturally understood in the framework of classical electrodynamics plus the concept of the spin magnetic moment. It is the force exerted on the spin moments of the carriers, in the system in which they are at rest, that contributes to both transverse imbalances. It can be concluded that, despite the validity of other approaches, the forces found in this article must be taken into account when trying to understand the effects produced by electric and spin currents.

**Acknowledgments**

The authors are indebted to M. A. García, P. Echenique, J. Hirsch, V. Golovach and A. Arnau for many fruitful discussions. This work has been supported by the Spanish Ministry of Innovation, Science and Technology and Spanish Ministry of Economy and Competitiveness through Research Projects MAT2015-67557-C2-1-P, MAT2017-86450-C4-1-R S2013/MIT-2850 NANOFRONTMAG, RTI2018-095856-B-C21, by the European Comission AMPHIBIAN (H2020-NMBP-2016-720853) and by the Comunidad de Madrid S2018/NMT-4321 NANOMAGCOST-CM

**Figures footnotes**

Figure 1. Thin metallic film with two currents of electrons with up and down spins running along it.

Figure 2. Spin Hall effect in a thin metallic film which contains a current (generated by a battery) running along its *y* direction.

Figure 3. Inverse Spin hall effect in a thin metallic film with a spin gradient between opposite *x* edges.

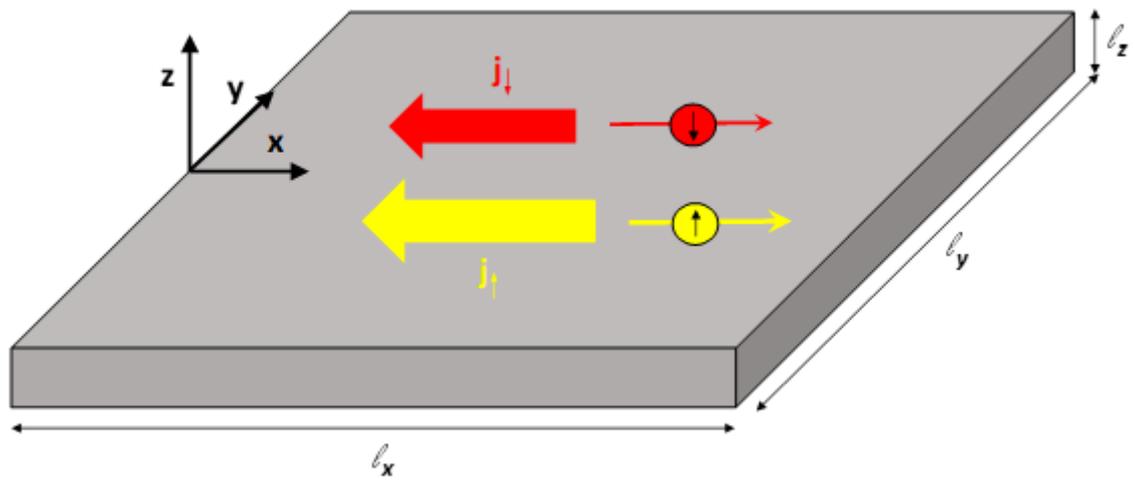

Figure 1

**Figure 2**

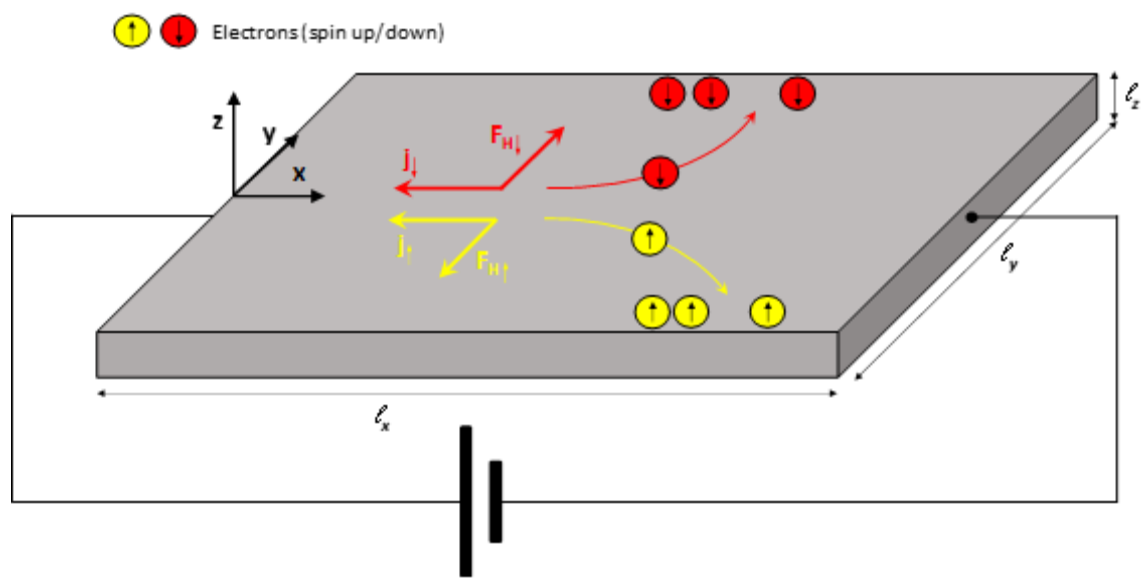

**Figure 3**

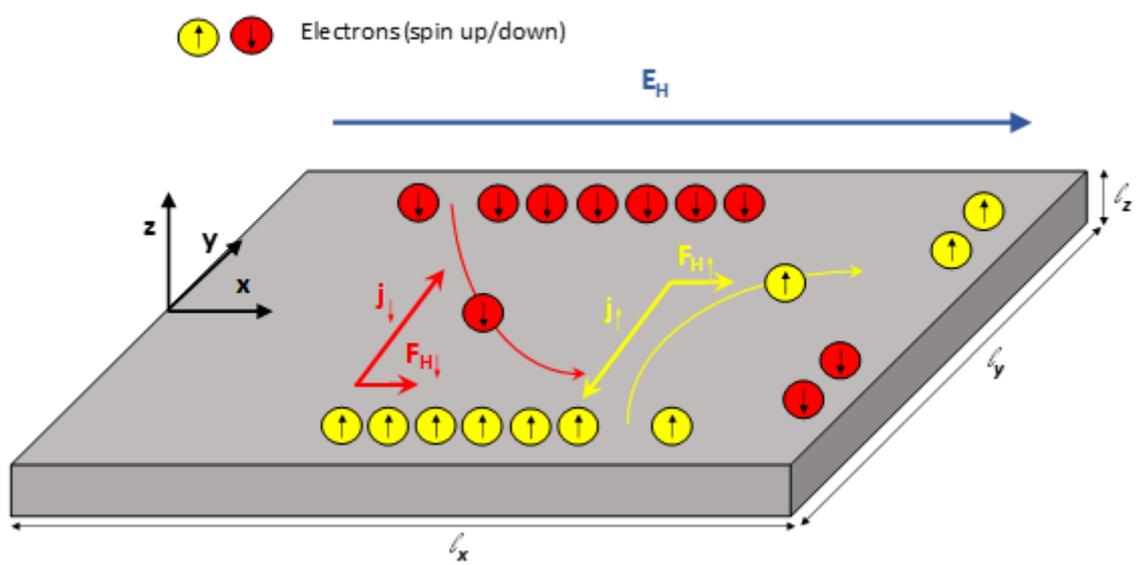

**Supplementary information**

**a)** Consider the case of a full spin polarized conduction band, thus, $j_y = j_{y\uparrow}$. Since the carriers current seen in $S_B F$ and the lattice current seen is $S_\uparrow F$ are the same, the inhomogeneous magnetic field produced by the lattice current in $S_\uparrow F$ is given by $B_z(x) = \alpha\mu_0 x j_y$ [1]. Where $\alpha$ is the geometrical coefficient defined as $\alpha = 8\frac{l_z}{l_y}$. The force acting on a spin up electron, that can be calculated in $S_\uparrow F$ becomes $f_{x\uparrow} = -\alpha\mu_0\mu_B j_y$. The net force on the whole sub-band is then $F_{x\uparrow} = -\alpha\mu_0 M_{s\uparrow} j_y$. Where $M_{s\uparrow} = n_\uparrow \mu_B = n\mu_B$

Now it can be calculated, also in $S_\uparrow F$, the force $F_{xB}$ on the background charge in which it moves with density current $j$. This force must be equal but opposite to $F_{x\uparrow}$ and proceeds from the Lorentz force undergone by the background current $j$, in the magnetic field $\alpha\mu_0 M_{s\uparrow}$, force given by $\alpha\mu_0 M_{s\uparrow} j_y$, that holds the required condition of being opposite to $F_{x\uparrow}$

In the text of the article, for the sake of clarity we have considered α=1 and have omitted the directions of forces, always along the *x-axis*, of the magnetization and field, also along the *z-axis* and the currents always, all of them, oriented along the *y axis*

**b)** The mutual force between two opposite spin sub-bands in relative motion, $j_\uparrow = n_\uparrow e v_\uparrow$ and $j_\downarrow = n_\downarrow e v_\downarrow$ can be obtained by calculating the Hall effect induced at each sub-band when an average field $B_z = \mu_0 M$ acts on them. This force acting on the electrons down becomes:

$$F_{\uparrow\downarrow} = j_y B_z = \mu_0 M (j_\downarrow - j_\uparrow) \quad (1')$$

Obviously, as required by the action and reaction principle, the force exerted on the spin up electrons is just the opposite to that acting on the spin down as can be immediately shown by inversing the sign of the relative density current $(j_\downarrow - j_\uparrow)$. It is important to indicate that the same forces should be obtained in any inertial reference framework. In fact, if we work on $S_\uparrow$ frame the force acting on the spin down assembly is immediately obtained as $\mu_0 M (j_\uparrow - j_\downarrow)$. However, in order to calculate the force acting on $S_\uparrow$ assembly one must consider that these electrons are at rest and that, consequently, the only force exerted on them must act on its spin magnetic moment [1].

**c)** The net force acting on the conduction band as depicted by eq. (5) was found assuming a spin depolarized background. When the magnetization of the lattice is *M\** the total force on the conduction band is given by

$$F^H = -\mu_0 M_s (j_\uparrow - j_\downarrow) + \mu_0 M^* (j_\uparrow + j_\downarrow) \quad (2')$$

as derived from eq. (5) after adding the net Lorentz force on both currents due to the lattice magnetization. This case is important in the analysis of the anomalous Hall effect.

Similarly, the net force inducing transverse spin separation can be found from eq. (7) to change by the effect of the lattice spin polarization to.

$$F^S = -\mu_0 M_s (j_\uparrow + j_\downarrow) + \mu_0 (M + M^*)(j_\uparrow - j_\downarrow) \quad (3')$$